\documentclass[12pt]{iopart}
\usepackage{graphicx}
\usepackage{bm}
\usepackage{times}
\usepackage[hypertex]{hyperref}
\usepackage{slashed}
\usepackage{epsf}
\usepackage{axodraw}
\def\br{\begin{eqnarray}}
\def\er{\end{eqnarray}}
\def\be{\begin{equation}}
\def\ee{\end{equation}}

\def\p{\phi}

\def\l{\lambda}

\def\m{\mu}

\def\({\left(}
\def\){\right)}
\def\s{\sigma}

\begin{document}

\title[Sterile neutrino dark matter, CDMS-II and a light Higgs boson]{Sterile neutrino dark matter, CDMS-II and a light Higgs boson}

\author{F. S. Queiroz, C. A. de S. Pires and P. S. Rodrigues da Silva}
\address{Departamento de F\'{\i}sica, Universidade Federal da
Para\'{\i}ba, \\ 
Caixa Postal 5008, 58051-970, Jo\~ao Pessoa - PB, Brazil.}

\ead{farinaldo@fisica.ufpb.br, cpires@fisica.ufpb.br and psilva@fisica.ufpb.br}

\date{\today}



\begin{abstract}
We add a singlet right handed neutrino plus a charged and a neutral singlet scalars to the standard model. This extension includes a discrete symmetry such that we obtain a heavy sterile neutrino which couples only to the electron and the new scalars. In this sense the singlet neutrino does not mix with ordinary ones and thus has no effect on Big Bang Nucleosynthesis. However, such sterile neutrino can be in equilibrium with electroweak particles in the early Universe due to its couplings to electrons and also because the Higgs boson mixes with the singlet scalars. We obtain that the sterile neutrino constitutes a dark matter candidate and analyze its direct detection in the light of current experiments. Our results show that if such a sterile neutrino is realized in nature, and CDMS-II experiment confirms its positive signal, dark matter demands a rather light Higgs boson with new Physics at some 500~GeV scale. 
\end{abstract}


\section{Introduction}\setcounter{equation}{0}
\label{intro}

The Dark Matter (DM) has become one of the most promising evidences in favor of Physics beyond standard model of electroweak interactions (SM). 
The converging paradigm to naturally explain this unseen component of matter is that it be a weakly interacting massive particle (WIMP), with mass scale ranging from tens of GeV to few TeV~\cite{DMmodels}.
With recent data from Wilkinson Microwave Anisotropy Probe (WMAP)~\cite{WMAP5}, and several DM direct detection experiments running and projected~\cite{cdms,xenon}, we have powerful new tools to investigate Particle Physics models which just provide new particle content that would explain this dark component. Besides, the Large Hadron Collider (LHC) is already running~\cite{LHC} and may be able to shed some light on the DM component too.

Another non-negotiable experimental evidence of new Physics is neutrino mass, once the SM does not provide a mechanism to explain this issue and we need to add some new ingredient to do this job. Even  better if this addition brings some information on other challenges not covered by SM.
It appears to us that these two clues about new Physics could have some common origin. Remember though that active neutrinos are underweight to merge into this picture, since their contribution to the density parameter, $\Omega_\nu$, is given by~\cite{zeldo},
\be
\Omega_\nu h^2 = \frac{\sum m_\nu}{93 {\mbox{eV}}}\,.
\label{active}
\ee

On the other hand, sterile neutrinos could possess the right features to accomplish the task of being dark and play some role in neutrino mass generation~\footnote{There are some interesting studies in literature where sterile neutrinos are at the keV scale, see for example Ref.~\cite{kusenko}, constituting a warm DM candidate. We are not going to pursue this possibility in this work.}. This is particularly true in the model of Ref.~\cite{GHSV}. There the authors managed to get a realistic model for neutrino mass by adding right handed neutrinos and scalar multiplets in such a way as to obtain small neutrino masses and provide a scenario for leptogenesis. The interesting thing is that one of these neutrinos is sterile and stable, and could be a DM candidate. Another such link between DM and sterile neutrinos is provided in Ref.~\cite{ma}, where a version of a left-right symmetric  model is extended with a $U(1)$ gauge symmetry. Using some non-supersymmetric R-parity symmetry the authors get a stable neutrino interacting with electron and a charged scalar and analyze its viability as DM candidate. In such model there is no direct detection constraint since the interaction is leptophilic. We can even imagine a third model where these effective interactions would appear, the so called $3-3-1$RH$_\nu$ model, a gauge extension of SM from $SU_L(2)\otimes U_Y(1)$ to $SU_L(3)\otimes U_N(1)$ with right handed neutrinos as part of a fundamental representation under $SU_L(3)$ for the leptons~\cite{valle,331rh}. Besides being capable of generating appropriate neutrino masses~\cite{331nm}, it has a rich spectrum of scalars and at least one studied case of a WIMP DM candidate~\cite{331wimp}, among other nice features that help to tackle unsolved problems of SM.

In this work we would like to link some features of these non-supersymmetric models~\cite{GHSV,ma,331rh} in a simpler effective low energy model for DM sterile neutrino. We suppose the presence of a right handed neutrino, which reveals itself to be sterile, and two singlet scalars, a neutral and a charged one. The role of these scalars is to give this neutrino a mass and to couple it with a charged lepton. We restrict this lepton to be the electron for simplicity, though it could be extended to muon and/or tau leptons. In this way, it could be seen as a low energy limit of the models discussed above~\cite{GHSV,ma,331rh}. We are not going to stick with none of these particular models, since the features we are going to study here can serve for each one in some specific regime, but it is important to know that some larger scheme can be introduced to justify our approach. Our main purpose in this work is to show that a sterile neutrino not only can represent the appropriate DM content of the Universe, in agreement with the latest results of CDMS-II (and XENON-10)~\cite{cdms,xenon}, but also it demands a light SM Higgs boson if DM has impinged a signal on the detector~\cite{kopp}. Maybe this is the only connection with electroweak physics, an unusual feature when we want to solve the neutrino mass puzzle, since neutrino mixing can be potentially dangerous exactly because of its effects on the electroweak side and are constrained by Big Bang Nucleosynthesis~\cite{BBN}. In this sense, while neutrino mass can be an indication of new Physics, it is the DM that can offer the source of detection of an entirely new branch of the leptonic sector not available for the standard interactions, although implying a constraint to Higgs boson mass.

The paper is organized as follows. In section~\ref{sec1}, we present our sterile neutrino model, identifying the physical fields (mass eigenstates) and impose the discrete symmetry which makes the sterile neutrino stable. Next, in section~\ref{sec2}, we compute the relic abundance of our WIMP candidate and its direct detection, paying some attention to the possible CDMS-II positive signal. We finally show our conclusions in section~\ref{sec3}.

\section{Sterile neutrino model}
\label{sec1}

The model consists of a small extension of SM by including a singlet Majorana neutrino, $N_R$, plus a singlet charged scalar, $\eta^\pm$ and a singlet neutral scalar, $\sigma$. Besides the mixing among the Higgs doublet and the new singlet scalars, there is an interaction term coupling these singlets to the leptons~\footnote{We will restrict our analysis to one family of leptons since this will be justified later when considering the embedding of this model in a realistic context considering neutrino mass.}, as can be seen in the following Lagrangian,
\br
{\cal L}\,\, & \supset & \,\, {\cal L}_{Kin} - \l_1\left(\overline{N_R^c} N_R\sigma + \overline{N_R} N_R^c \sigma^{*}\right) + \l_2\left(\overline{N_R^c} e_R\eta^+ + \overline{e_R} N_R^c \eta^-\right) 
\nonumber \\
&&- V^\prime(\phi\,,\eta^\pm\,,\sigma)  \,,
\label{lagr}
\er
where ${\cal L}_{Kin}$ is the Lagrangian term involving the kinetic terms for the singlet neutrino as well as for the singlet scalars, including the $\eta^\pm$ interaction with SM gauge bosons, $Z$ and $\gamma$. From this Lagrangian we observe that the scalars carry two units of lepton number (called bileptons) if we admit lepton number conservation. Notice also that $\eta^\pm$ will have interaction terms with the photon and $Z^0$ boson, since it possess nonzero hypercharge.  The scalar potential in the above Lagrangian, $V^\prime(\phi\,,\eta^\pm\,,\sigma)$, embraces the new terms to be added to the SM Higgs potential, in order to form the complete scalar potential, $V(\phi\,,\eta^\pm\,,\sigma)=V^\prime(\phi\,,\eta^\pm\,,\sigma)+V_{SM}$, given by,
\br
&&V(\phi\,,\eta^\pm\,,\sigma) = m_\phi^2\phi^\dagger\phi + m_\s^2\s^{*}\s + \m_\eta^2 \eta^+ \eta^- + \l_\phi(\phi^\dagger\phi)^2 + \l_\s(\s^{*}\s)^2
\nonumber \\
&&+ \l_\eta(\eta^+\eta^-)^2 + \l_3(\phi^\dagger\phi)(\eta^+\eta^-) + \l_4\eta^+\eta^-\s^{*}\s
+\l_5\phi^\dagger\phi\s^{*}\s\,.
\label{pot}
\er
where the field $\phi$ represents the usual SM scalar $SU_L(2)$ doublet. 

Observe that our Lagrangian, Eq.~(\ref{lagr}), allows the presence of some extra terms not included here. For example, we could add a term connecting the lepton doublet, $L\equiv (\nu_{eL}\,,e_L)^T$, with the singlet neutrino through the SM scalar doublet,
$\bar{L}\phi N_R + h.c.$, however, we want to avoid such terms since we are interested in explaining the DM component as constituted by the singlet neutrino. This term would imply its mixing with standard left handed neutrinos and consequently its decay. We choose to get rid of these kind of couplings by assigning a discrete symmetry to $N_R$ and $\eta^\pm$, namely,
\be
(N_R\,,\eta^+)\rightarrow (-N_R\,,-\eta^+)\,.
\label{discrete}
\ee
That is what make this singlet neutrino a sterile one, since it does not couple to any gauge boson of the SM and would leave no sign of its existence through the known interactions. This model is what we call a leptophilic model for the charged scalar  in the sense that it does not mediate any interaction involving quarks. Even if we had enforced this scalar to couple to some quark, this one should be an exotic quark, meaning that it would carry lepton number and be a bilepton too, if we want to preserve lepton symmetry from being broken explicitly. This kind of particles, scalar and quark bileptons, are very common in the class of models called 3-3-1~\cite{331rh,vicente} and a neutral scalar bilepton was already pointed as a WIMP DM candidate in a version of 3-3-1 with right-handed neutrinos in the fundamental triplet representation~\cite{331wimp}.

In this work we rely upon the possibility that some high energy fundamental theory is going to properly generate this effective model. In this way we consider the charged scalar mass scale as a free parameter, which can be of the order of the electroweak symmetry breaking scale. Meanwhile, the neutrino mass can be generated by spontaneous symmetry breaking due to a nonzero vacuum expectation value (VEV) acquired by $\sigma$, $v_\s$, which we suppose to be real and assuming values around TeV scale,
\be
\s = \frac{1}{\sqrt{2}}(v_\s + \s_R +i\s_I)\,.
\label{vev}
\ee
A Majoron should emerge from this breaking mechanism, but it turns out to be a safe Majoron since it does not couple to the electroweak sector except for the Higgs, putting us in a comfortable situation to not worry about its presence in the spectrum. Then, from Eq.~(\ref{lagr}), we get a Majorana neutrino mass that can range from tens of GeV to hundreds of GeV,
\be
M_N = \l_1 v_\sigma\sqrt{2}\,.
\label{neutmass}
\ee

Issues concerning light neutrino masses can be addressed by different approaches developed in Refs.\cite{GHSV,ma,331rh}, which seem to be particularly interesting to embed our model. We noticed that our right-handed neutrino has exactly the couplings with a neutral singlet scalar and/or the charged singlet scalar in their models. The models in Refs.~\cite{GHSV,ma} have already proposed that there is a stable sterile neutrino in their spectra, although they did not deeply surveyed this possibility as we are going to do here. In the case of the model in Ref.~\cite{ma}, that we can recover when $\s$ decouples, the authors had a limitation that their neutrino does not give a sign in direct detection experiments. Also, for the case where the charged scalar is almost degenerate in mass with the neutrino, it is fair to mention that coannihilation plays an important role, changing a little their quantitative results, though it do not really change their qualitative aspects. As for the case of the model in Ref.~\cite{GHSV}, a complete analysis of abundance and direct detection is lacking. Even if we cannot directly compare our results, it remains as a good reference point for the  structure of our model, since our neutrino is much
like their third neutrino, it can be a plausible DM candidate, a WIMP. In fact, our proposed charged current involves a sterile neutrino and the electron, instead of a heavier lepton. We could impose a coupling to a heavier charged lepton, muon or tau, but we not expect big quantitative differences concerning these alternative choices though, and will stick with the coupling with electron only, recovering the main aspects of their DM sector when $\eta^\pm$ decouples.
In what concerns the 3-3-1 model~\cite{valle,331rh}, the possibility of a sterile neutrino as DM candidate is still under study and again we have no direct way to contrast results.

Considering the limits imposed by LEP on the direct production of such heavy neutrino and the charged scalar bilepton, we can consider the sterile neutrino mass, a Majorana one, above 46~GeV, and the charged scalar mass above $100$~GeV~\cite{PDG}. In this model the Yukawa couplings can be naturally of order one, and we will explore this fact also imposing that the Higgs boson mass be above the experimental limit $M_H \geq 114.4$~GeV~\cite{PDG} and less than $300$~GeV~\footnote{An exception to the upper bound on the Higgs mass will be taken when discussing small sterile neutrino mass to account for recent CDMS-II results.}. 

\subsection{Scalar mass spectrum}

The scalar masses can be obtained by diagonalizing the mass matrix coming from the new terms in the potential Eq.~(\ref{pot}) and the Higgs potential in SM. We take the basis for the real part of neutral scalars, $R_\phi$ and $R_\sigma$, resulting in the following mass matrix:
%
%

%
\br
M^2 = \left(
\begin{array}{cc}
\l_\phi v_\phi^2 & \frac{\l_5}{2}v_\s v_\phi \\
\frac{\l_5}{2}v_\s v_\phi & \l_\s v_\s^2	
\end{array} \right)\,,
\label{mm}
\er
which was obtained after using the minimum conditions for the scalar potential,
\br
\mu_\s^2 + \l_\s v_\s^2+\frac{\l_5}{2}v_\phi^2 &=& 0\,, \nonumber \\
\mu_\phi^2 + \l_\phi v_\phi^2 +\frac{\l_5}{2}v_\s^2 &=& 0\,.
\label{minimum}
\er
As for the pseudoscalars, those coming from the Higgs doublet play the same role as before, providing the usual Goldstone bosons for the SM, while the new pseudoscalar coming from the singlet neutral scalar, becomes a massless Majoron, $J$, which does not couple to the electroweak sector except through the standard Higgs boson. We keep it massless throughout this work, though it could get a mass from higher order terms or radiative corrections and still be a light particle, not changing our results. Of course it can be a DM candidate too, but a non-thermal one, and we are not going to analyze this possibility here.

The mass matrix, Eq.~(\ref{mm}), can be easily diagonalized to give the following mass eigenvalues,
\br
M_H^2 &=& \l_\s v_\s^2 + \l_\phi v_\phi^2 - \sqrt{\l_\s^2 v_\s^4 +\l_5^2 v_\p^2 v_\s^2 -2\l_\s \l_\p v_\p^2 v_\s^2 + \l_\p^2 v_\p^4}\,,
\nonumber \\
M_S^2 &=& \l_\s v_\s^2 + \l_\phi v_\phi^2 + \sqrt{\l_\s^2 v_\s^4 +\l_5^2 v_\p^2 v_\s^2 -2\l_\s \l_\p v_\p^2 v_\s^2 + \l_\p^2 v_\p^4}\,,
\label{meig}
\er
with the respective approximate mass eigenstates, the Higgs and the new neutral scalar,
\br
H &=& \frac{1}{1+\left(\frac{\l_5 v_\p}{2\l_\s v_\s}\right)^2}\left(R_\p - \frac{\l_5 v_\p}{2\l_\s v_\s} R_\s \right)\,, \nonumber \\
S &=& -\frac{1}{1+\left(\frac{\l_5 v_\p}{2\l_\s v_\s}\right)^2}\left(R_\s + \frac{\l_5 v_\p}{2\l_\s v_\s} R_\p \right)\,.
\label{meigs}
\er

Finally, the charged scalar bilepton, $\eta^\pm$, has a mass term given by,
\be
m_\eta^2 = \m_\eta^2 + \frac{\l_3}{2}v_\p^2\,,
\label{cscalar}
\ee
which can be considered as a free parameter since $\m_\eta$ is unknown and can assume any value. We will always take this mass value bigger than the sterile neutrino mass in order to guarantee the neutrino stability. Next we analyze the viability of the sterile neutrino as a WIMP DM candidate by computing its relic abundance and direct detection signals.

\section{WIMP relic abundance and direct detection}
\label{sec2}

Based on the discussion in Sec.~\ref{intro}, the neutral stable particle of this model which plays the role of a WIMP is the heavy singlet Majorana neutrino, the sterile neutrino. Its stability is guaranteed by a discrete symmetry (the $Z_2$ symmetry mentioned above) which avoids its coupling to left handed neutrinos, while its mass is less than the new charged singlet scalar.
However, it must reproduce the right relic abundance, $\Omega_{DM}\simeq 22\%$, and at the same time it cannot conflict with established direct detection experiments~\cite{cdms,xenon}. Let us first consider the computation of its relic density.

We depart from the assumption that the sterile neutrino was in thermal equilibrium with the SM particles until its decoupling from the thermal bath, the freeze-out epoch. This thermal equilibrium is possible thanks to the interaction among the sterile neutrino and the electron and also to the Higgs boson, through the mixing with the neutral scalar singlet. Since we are talking about a WIMP, the freeze-out temperature is reached when it is non-relativistic, meaning that the energy scale to be considered is about $M_N/20$, where $M_N$ is the sterile neutrino mass.   
We can then follow the evolution of the particle number density, $n$, through the Boltzmann equation while the Universe undergoes its expansion,
\be
\frac{dn}{dt}+3Hn = -<\sigma v_{r}>(n^2-n^2_{eq}),
\label{boltzmann}
\ee
where, $H$ is the Hubble parameter, which can be written as $H^2=8\pi\rho/3M_{Pl}^2$ for a flat Universe, $<\sigma v_{r}>$ is the WIMP annihilation cross section thermally averaged with the relative velocity, $v_{r}$ and $n_{eq}$ is the particle number density at equilibrium. In the expression for the Hubble parameter, $M_{Pl}\approx 1\times 10^{19}$~GeV, which is the Planck scale and $\rho$ is the energy density of the Universe. Since WIMPs are intrinsically non-relativistic at the time of decoupling ($T<<M_{N}$ in our case), its equilibrium number density is given by the quantum statistical mechanics expression,
\be
n_{eq}=g_N\left(\frac{M_{N}T}{2\pi}\right)^\frac{3}{2}e^{-\frac{M_{N}}{T}},
\label{nequilibrium}
\ee
where $g_N$ is the number of degrees of freedom of the WIMP. In general, we follow the standard procedure derived in Ref.~\cite{DMmodels,SWO} by defining $x\equiv M_{N}/T$ and using the non-relativistic approximation for the squared center of momentum energy, $s=4M_N^2+M_N^2v_r^2$, to expand the thermally averaged cross section till the first power in $v_r^2$, which allows us to write it as,
\be
<\sigma v_{r}>\approx a+\frac{6b}{x},
\label{thermalexpansion}
\ee
where $a$ and $b$ are the model dependent parameters. Then we can write the relic WIMP abundance as,
\be
\Omega_N h^2 \approx \frac{1.04\times10^9}{M_{Pl}}\frac{x_F}{\sqrt{g^*}(a+\frac{3b}{x_F})}\,,
\label{omega}
\ee
where $x_F$ is $x\equiv M_{N}/T$ computed at the WIMP decoupling temperature given by,
\be
x_F = \ln{\left[c(c+2)\sqrt{\frac{45}{8}}\frac{g_\phi}{2\pi^3}\, \frac{M_N\, M_{Pl}\, (a+\frac{6b}{x_F})}{\sqrt{g^* x_F}}\right]}\,,
\label{xf}
\ee
which, as we mentioned before, is generally close to $x_F\approx 20$.
In this expression, $g^*$ is the number of degrees of freedom of all relativistic SM particles in thermal equilibrium with the WIMP at freeze-out temperature, which can be easily computed to give $g^*=106.75$, with  $g_N=2$, while the constant $c$ is taken to be of order of one. The exact value of this constant is unimportant since it has only a small effect in the logarithmic dependence of $x_F$.

\begin{figure}[h]
\centering
{} \qquad\allowbreak
\begin{picture}(95,59)(0,0)
\Text(15.0,50.0)[r]{$N_R$}
\Line(16.0,50.0)(58.0,50.0) 
\Text(80.0,50.0)[l]{$\bar{e}$}
\ArrowLine(79.0,50.0)(58.0,50.0) 
\Text(54.0,40.0)[r]{$\eta^-$}
\DashArrowLine(58.0,50.0)(58.0,30.0){1.0} 
\Text(15.0,30.0)[r]{$N_R$}
\Line(16.0,30.0)(58.0,30.0) 
\Text(80.0,30.0)[l]{$e$}
\ArrowLine(58.0,30.0)(79.0,30.0) 
\end{picture} \ 
\begin{picture}(95,59)(0,0)
\Text(15.0,50.0)[r]{$N_R$}
\Line(16.0,50.0)(37.0,40.0) 
\Text(15.0,30.0)[r]{$N_R$}
\Line(16.0,30.0)(37.0,40.0) 
\Text(47.0,41.0)[b]{$S$,$H$}
\DashLine(37.0,40.0)(58.0,40.0){1.0}
\Text(80.0,50.0)[l]{$B$}
\Photon(58.0,40.0)(79.0,50.0){3.0}{3} 
\Text(80.0,30.0)[l]{$B$}
\Photon(79.0,30.0)(58.0,40.0){3.0}{3}
\end{picture} \ 
\begin{picture}(95,59)(0,0)
\Text(15.0,50.0)[r]{$N_R$}
\Line(16.0,50.0)(37.0,40.0) 
\Text(15.0,30.0)[r]{$N_R$}
\Line(16.0,30.0)(37.0,40.0) 
\Text(47.0,41.0)[b]{$H$,$S$,$J$}
\DashLine(37.0,40.0)(58.0,40.0){1.0}
\Text(80.0,50.0)[l]{$S_1$}
\DashLine(58.0,40.0)(79.0,50.0){1.0}
\Text(80.0,30.0)[l]{$S_2$}
\DashLine(58.0,40.0)(79.0,30.0){1.0}
\end{picture} \ 
{} \qquad\allowbreak
\begin{picture}(95,59)(0,0)
\Text(15.0,50.0)[r]{$N_R$}
\Line(16.0,50.0)(58.0,50.0) 
\Text(80.0,50.0)[l]{$S_1$}
\DashLine(58.0,50.0)(79.0,50.0){1.0}
\Text(57.0,40.0)[r]{$N_R$}
\Line(58.0,50.0)(58.0,30.0) 
\Text(15.0,30.0)[r]{$N_R$}
\Line(16.0,30.0)(58.0,30.0) 
\Text(80.0,30.0)[l]{$S_2$}
\DashLine(58.0,30.0)(79.0,30.0){1.0}
\end{picture} \
\begin{picture}(95,59)(0,0)
\Text(15.0,50.0)[r]{$N_R$}
\Line(16.0,50.0)(37.0,40.0) 
\Text(15.0,30.0)[r]{$N_R$}
\Line(16.0,30.0)(37.0,40.0) 
\Text(47.0,41.0)[b]{$H$,$S$}
\DashLine(37.0,40.0)(58.0,40.0){1.0}
\Text(80.0,50.0)[l]{$f$}
\ArrowLine(58.0,40.0)(79.0,50.0) 
\Text(80.0,30.0)[l]{$\bar{f}$}
\ArrowLine(79.0,30.0)(58.0,40.0) 
\end{picture} \ 
\caption{Sterile neutrino annihilation channels that contribute to $\Omega_{N}$. In these graphs, $B$ stands for a $W^\pm$ or a $Z$ gauge boson, $S_1$ and $S_2$ represent any possible combination of scalars in the model, $H$, $S$ and $J$, and $f$ is a fermion.}
\label{fig:df1}
\end{figure}
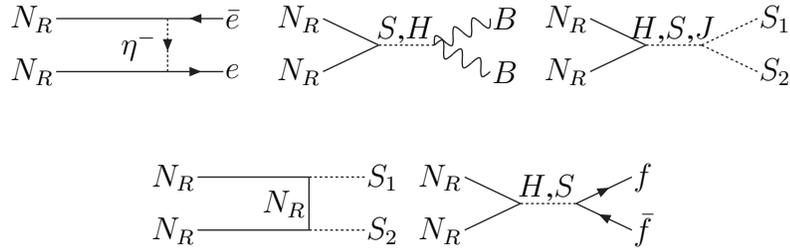

In a first naive approximation we could solve Eq.~\ref{xf} iteratively and plug the result in Eq.~\ref{omega}, by considering the annihilation channel $N+N\rightarrow X + Y$, where $X$ and $Y$ can be any SM particle or even the light singlet scalars of the model (see figure~\ref{fig:df1}). 

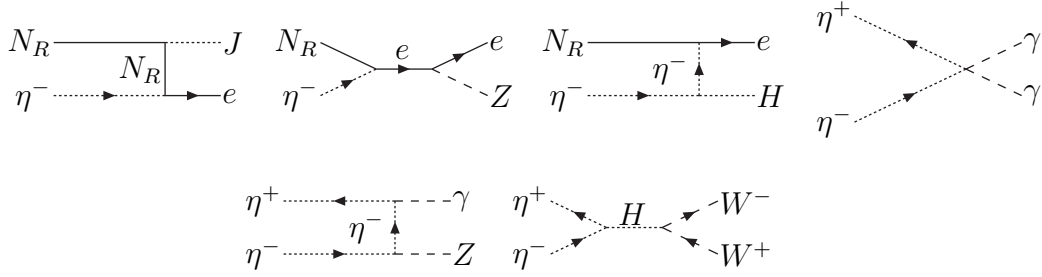
\begin{figure}[h]
\centering
{} \qquad\allowbreak
\begin{picture}(95,59)(0,0)
\Text(15.0,50.0)[r]{$N_R$}
\Line(16.0,50.0)(58.0,50.0) 
\Text(80.0,50.0)[l]{$J$}
\DashLine(58.0,50.0)(79.0,50.0){1.0}
\Text(57.0,40.0)[r]{$N_R$}
\Line(58.0,50.0)(58.0,30.0) 
\Text(15.0,30.0)[r]{$\eta^-$}
\DashArrowLine(16.0,30.0)(58.0,30.0){1.0} 
\Text(80.0,30.0)[l]{$e$}
\ArrowLine(58.0,30.0)(79.0,30.0) 
\end{picture} \ 
\begin{picture}(95,59)(0,0)
\Text(15.0,50.0)[r]{$N_R$}
\Line(16.0,50.0)(37.0,40.0) 
\Text(15.0,30.0)[r]{$\eta^-$}
\DashArrowLine(16.0,30.0)(37.0,40.0){1.0} 
\Text(47.0,44.0)[b]{$e$}
\ArrowLine(37.0,40.0)(58.0,40.0) 
\Text(80.0,50.0)[l]{$e$}
\ArrowLine(58.0,40.0)(79.0,50.0) 
\Text(80.0,30.0)[l]{$Z$}
\DashLine(58.0,40.0)(79.0,30.0){3.0} 
\end{picture} \ 
\begin{picture}(95,59)(0,0)
\Text(15.0,50.0)[r]{$N_R$}
\Line(16.0,50.0)(58.0,50.0) 
\Text(80.0,50.0)[l]{$e$}
\ArrowLine(58.0,50.0)(79.0,50.0) 
\Text(54.0,40.0)[r]{$\eta^-$}
\DashArrowLine(58.0,30.0)(58.0,50.0){1.0} 
\Text(15.0,30.0)[r]{$\eta^-$}
\DashArrowLine(16.0,30.0)(58.0,30.0){1.0} 
\Text(80.0,30.0)[l]{$H$}
\DashLine(58.0,30.0)(79.0,30.0){1.0}
\end{picture} \ 
\begin{picture}(95,59)(0,0)
\Text(15.0,60.0)[r]{$\eta^+$}
\DashArrowLine(58.0,40.0)(16.0,60.0){1.0} 
\Text(15.0,20.0)[r]{$\eta^-$}
\DashArrowLine(16.0,20.0)(58.0,40.0){1.0} 
\Text(80.0,50.0)[l]{$\gamma$}
\DashLine(58.0,40.0)(79.0,50.0){3.0} 
\Text(80.0,30.0)[l]{$\gamma$}
\DashLine(58.0,40.0)(79.0,30.0){3.0} 
\end{picture} \ 
\begin{picture}(95,59)(0,0)
\Text(15.0,50.0)[r]{$\eta^+$}
\DashArrowLine(58.0,50.0)(16.0,50.0){1.0} 
\Text(80.0,50.0)[l]{$\gamma$}
\DashLine(58.0,50.0)(79.0,50.0){3.0} 
\Text(54.0,40.0)[r]{$\eta^-$}
\DashArrowLine(58.0,30.0)(58.0,50.0){1.0} 
\Text(15.0,30.0)[r]{$\eta^-$}
\DashArrowLine(16.0,30.0)(58.0,30.0){1.0} 
\Text(80.0,30.0)[l]{$Z$}
\DashLine(58.0,30.0)(79.0,30.0){3.0} 
\end{picture} \ 
\begin{picture}(95,59)(0,0)
\Text(15.0,50.0)[r]{$\eta^+$}
\DashArrowLine(37.0,40.0)(16.0,50.0){1.0} 
\Text(15.0,30.0)[r]{$\eta^-$}
\DashArrowLine(16.0,30.0)(37.0,40.0){1.0} 
\Text(47.0,41.0)[b]{$H$}
\DashLine(37.0,40.0)(58.0,40.0){1.0}
\Text(80.0,50.0)[l]{$W^-$}
\DashArrowLine(58.0,40.0)(79.0,50.0){3.0} 
\Text(80.0,30.0)[l]{$W^+$}
\DashArrowLine(79.0,30.0)(58.0,40.0){3.0} 
\end{picture} \ 
\caption{Some coannihilation channels that contribute to $\Omega_{N}$.}
\label{fig:df2}
\end{figure}

However, it has to be kept in mind that such a procedure, while giving a rough estimate of $\Omega_N$, would not be appropriate for all regions of the available parameter space. Many coannihilation channels would be important when $M_\eta\simeq M_N$ (see some of the coannihilation contributions in figure~\ref{fig:df2}). As we can  be checked in a more accurate treatment of the problem, even a $100$~GeV mass difference can lead to significant contributions to $\Omega_N$ from these channels for some region of the coupling constants. For this reason, our approach in solving Boltzmann equation cannot rely on the simplified form given by Eq.~(\ref{boltzmann}), instead it will be carried out by using the micrOMEGAs package~\cite{micromegas}, which does the job allowing for the possibility of accurately taking the coannihilation channels into account. The Feynman rules used in micrOMEGAs were generated using the LanHEP package~\cite{lanhep}.

We will also rely upon micrOMEGAs in order to compute the direct detection cross section per nucleon and check the model through the recent bounds imposed by XENON~\cite{xenon} and CDMS-II~\cite{cdms} experiments. Basically, a DM particle can scatter with a nucleon thanks to its coupling to quarks via some SM mediator, leaving its trace through the observed recoil of the nuclei in the dedicated experiments~\cite{DMmodels,detdirReview}.

The sterile neutrino couples to quarks only through Higgs exchange and we have a spin independent (SI) scattering. It is useful to define the following average amplitude:
\be
\langle {\cal M}\rangle = i \alpha_q \langle {\bar q} q\rangle\,,
\label{ampli2}
\ee
where $\alpha_q$ is the effective quartic coupling of the sterile neutrinos with quarks,
\br
\alpha_q &=& -\left(G_F\sqrt{2}\right)^\frac{1}{2}\frac{m_q}{M_H^2}\l_{N-N-H}\,.
\label{alphaq}
\er
with,
\be
\l_{N-N-H} = \l_1 \l_5 \frac{\sqrt{2} v}{(4 \l_\s v_\s)}\,.
\label{coupl}
\ee
The matrix elements for the quarks in the nucleon can be written as follows~\cite{DMmodels},
\br
\langle {\bar q} q\rangle &=& \frac{m_{p,n}}{m_q}f_{T_q}^{p,n}\,\,\,\,\,\,\,\,\,\,\,\,\,\,\mbox{ for  u, d, s}\,, \nonumber \\
\langle {\bar q} q\rangle &=& \frac{2}{27}\frac{m_{p,n}}{m_q}f_{T_G}^{p,n}\,\,\,\,\,\,\,\mbox{ for  c, b, t}\,.
\label{quarksnucl}
\er
Hereafter the subscripts $p$ and $n$ label the proton and the neutron, respectively.

In order to obtain the WIMP-nucleon coupling it remains to sum over the quarks. We get,
\be
f_{p,n} = m_{p,n}\sum_{q=u,d,s}\frac{\alpha_q}{m_q}f_{T_q}^{p,n} +\frac{2}{27}m_{p,n}f_{T_G}^{p,n}\sum_{q=c,b,t}\frac{\alpha_q}{m_q}\,,
\label{wimpnucl}
\ee
where $f_{T_u}^p=0.023$, $f_{T_d}^p=0.033$, $f_{T_s}^p=0.26$, $f_{T_u}^n=0.018$, $f_{T_d}^n=0.042$, $f_{T_s}^n=0.26$ (see Ref.~\cite{micromegas}) and $f_{T_G}^{p,n}$, which is due to the neutrino coupling to gluons through loops of heavy quarks, is obtained from, 
\be
f_{T_G}^{p,n}=1-\sum_{q=u,d,s}f_{T_q}^{p,n}\,,
\ee
It is possible to write WIMP-nucleus cross section after summing the nucleons in the target, yielding, at zero momentum transfer~\cite{DMmodels,micromegas},
\be
\sigma_0 = \frac{4 \mu_r^2}{\pi}\left( Z f_p + (A-Z) f_n \right)^2\,,
\label{wimpnucleo}
\ee
where $\mu_r=M_N m_A/(M_N + m_A)$ is the reduced WIMP mass,
with, $m_A$ the nucleus mass, $Z$ the atomic number and $A$ the atomic mass.
Usually, the constraints on WIMP-nuclei elastic experiments is expressed in terms of the WIMP-nucleon cross section, which in the SI case reads,
\be
\sigma_{p,n}^{SI} = \sigma_0 \frac{m_{p,n}^2}{\mu_r^2 A^2}\,.
\label{wimpnucleoncs}
\ee

Next we present our results for the relic abundance and direct detection for some regions of the parameter space considering two different values of Higgs mass and also the whole range of masses between $114.4 \leq M_H \leq 300$~GeV. We finally comment on the latest CDMS-II~\cite{cdms} results which point to a favored region for the DM, although these results cannot yet be considered as a clear signal of detection when faced to its background.

\subsection{Relic Abundance}
\label{abundance}

In order to estimate the contribution of the sterile neutrino to the relic abundance of DM in the Universe, it is necessary to make some assumptions on the parameter space of the model.
This sterile neutrino model have quite a number of free parameters besides those of the SM. There is one VEV for the neutral singlet scalar, $v_\s$, which we suppose to be larger or equal 500~GeV since it has to be related to new physics at a high energy scale, eight coupling constants from the fermionic sector and the scalar potential, and the charged singlet scalar mass. If we demand that this should be a naturally motivated model, without much fine tunning, all the couplings are to be of order one. Also, we want the model to be testable at LHC, so we impose that mass scales are at most 1~TeV. 
Concerning the experimental results for the dark matter relic abundance, we consider in this work the WMAP 5-year alternative with $95\%$CL~\cite{WMAP5}, namely $0.108\leq \Omega_{DM} h^2 \leq 0.121$. 
With these basic assumptions we have made some quantitative predictions for the relic density of this sterile neutrino in some specific regimes of the model proposed above. 

\begin{figure}[h]
\centering
\includegraphics[width=0.45\columnwidth]{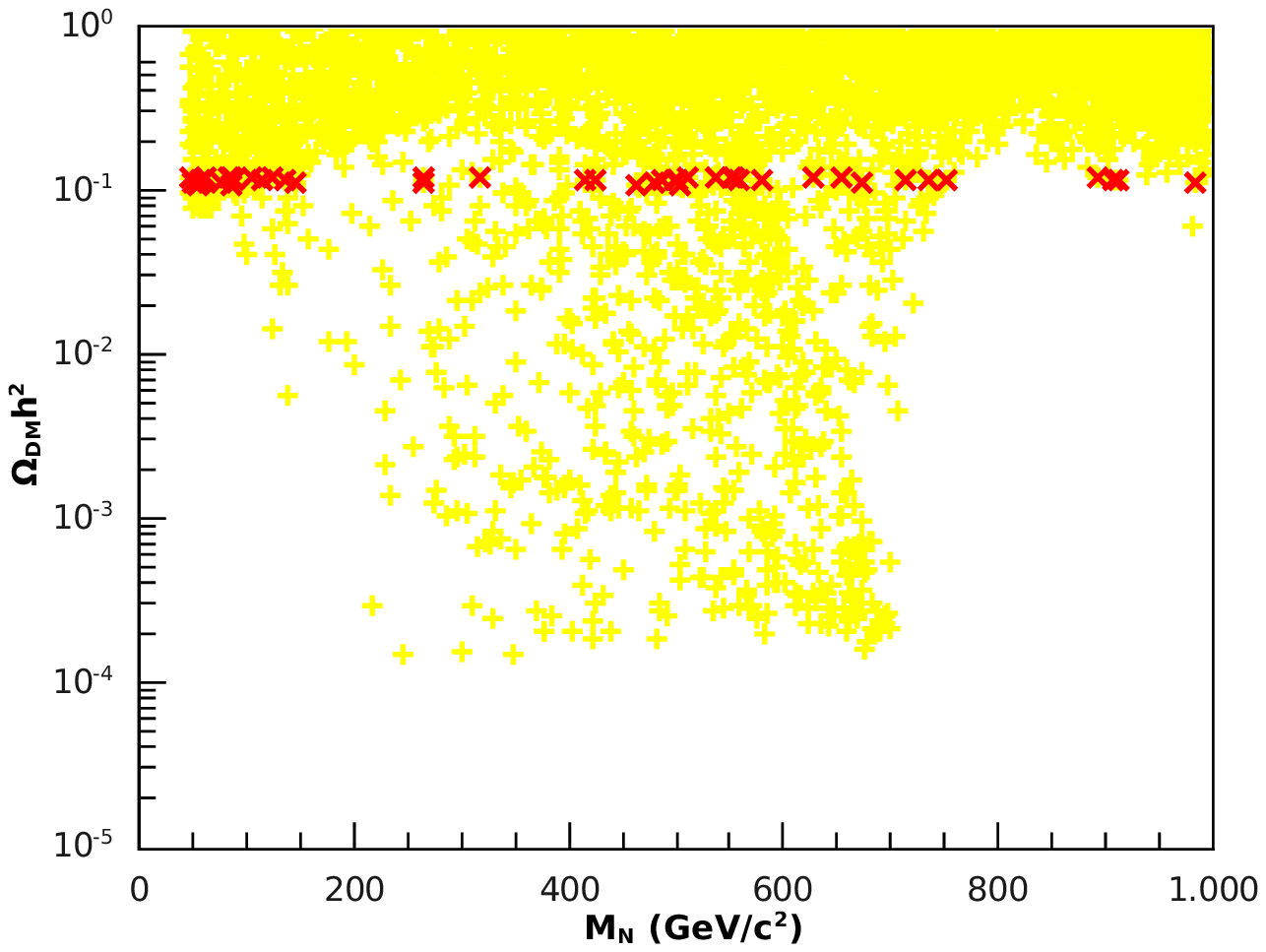}
\includegraphics[width=0.45\columnwidth]{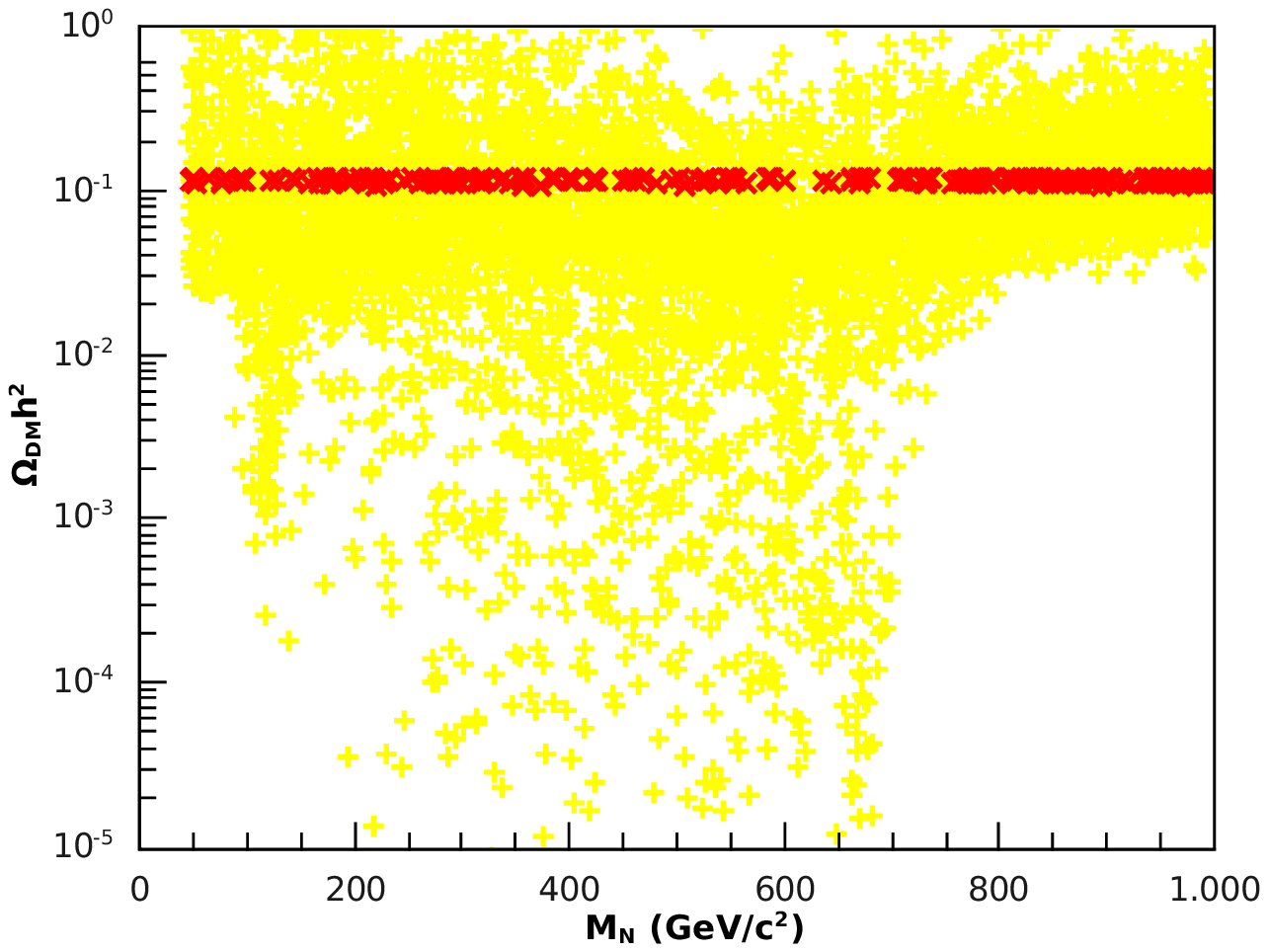}
\caption{Scatter plot for the relic abundance as a function of sterile neutrino mass. The left panel was computed  without coannihilation, while the right panel includes coannihilation. In these graphs the yellow cross represents points for several possible outcomes for the model considered, while the red cross represents only those points in parameter space which are in agreement with WMAP 5-year run within $95\%$ CL.}
\label{fig:1}
\end{figure}

\begin{figure}[h]
\centering
\includegraphics[width=0.45\columnwidth]{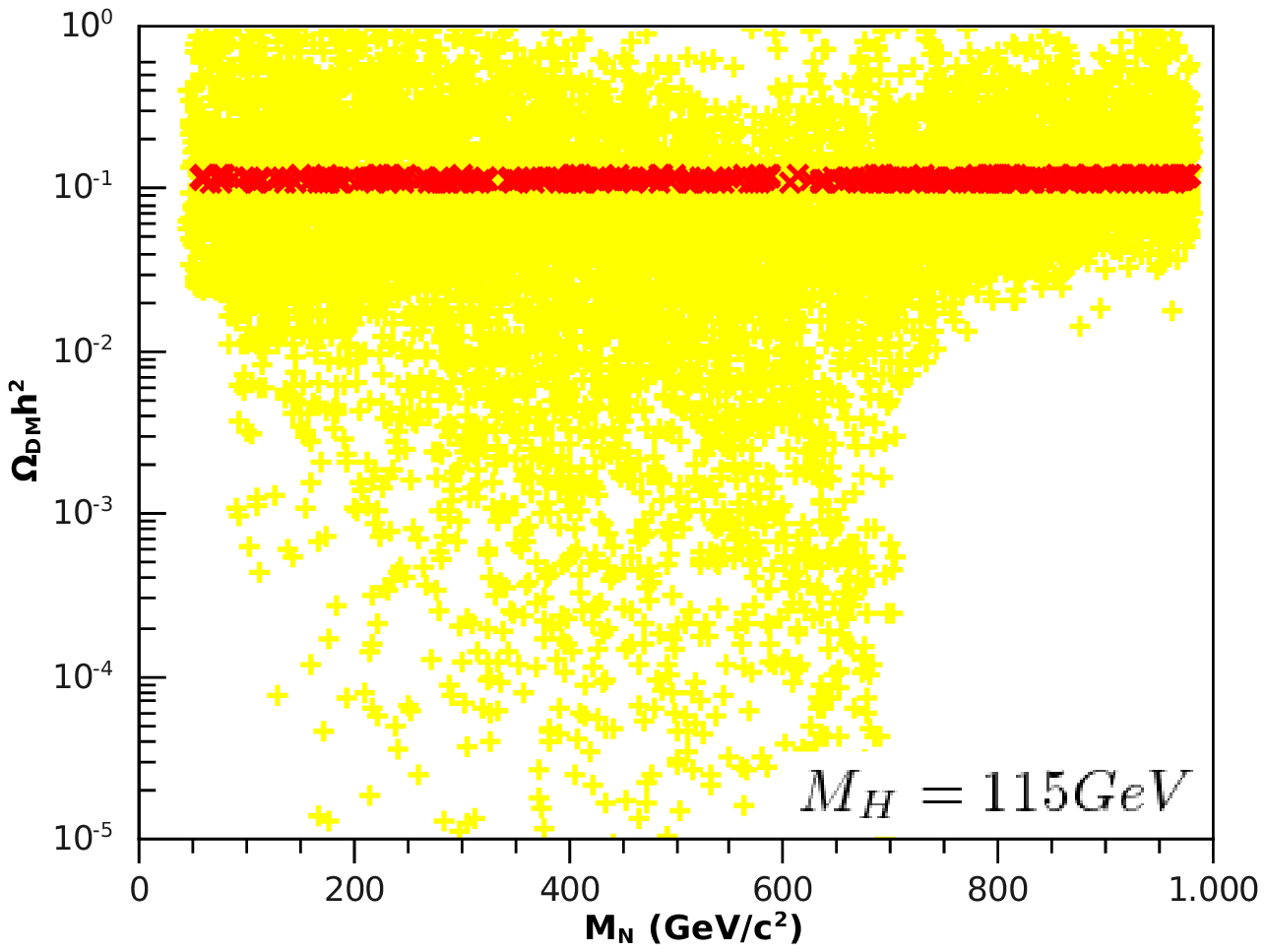}
\includegraphics[width=0.45\columnwidth]{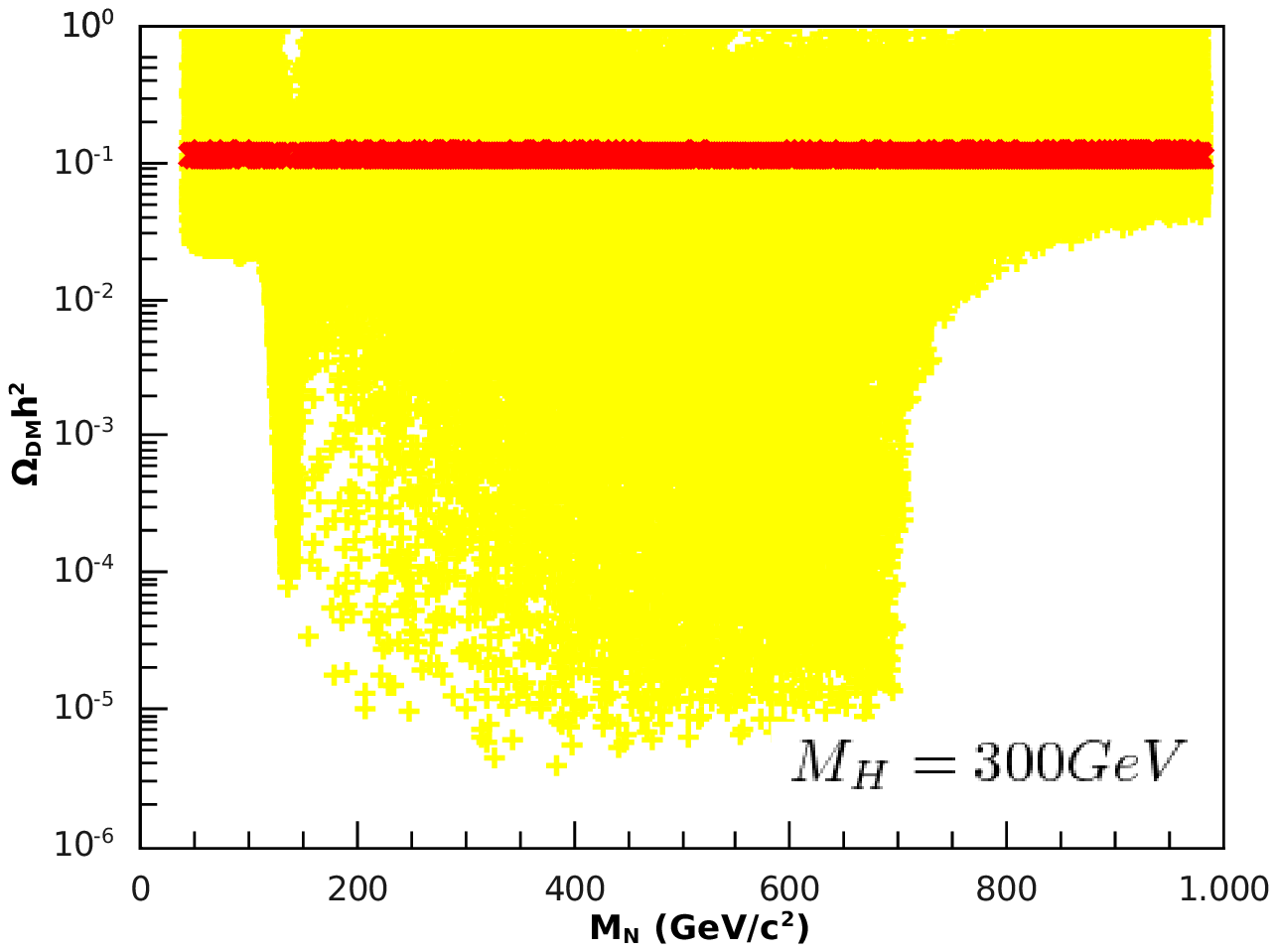}
\caption{Scatter plot for the relic abundance as a function of sterile neutrino mass for Higgs mass of 115~GeV (left panel) and 300~GeV (right panel), with coannihilation. The yellow cross represents points for several possible outcomes for the model considered, while the red cross represents only those points in parameter space which are in agreement with WMAP 5-year run within $95\%$ CL.}
\label{fig:2}
\end{figure}

We first compute the relic density for several points in the parameter space, for couplings varying about 0.1, a neutrino mass between $M_N=46$~GeV and $M_N=1$~TeV, and a range of Higgs mass between $M_H=114.4$~GeV to $M_H=300$~GeV. We consider two cases, one without coannihilation with $\eta^\pm$ and another including the coannihilation. For the first case we suppose a mass difference between the neutrino and charged scalar of 100~GeV. In the second case, such a difference is only of order of 10~GeV~\footnote{The only situation when this difference in mass is not 10~GeV is when we have neutrino masses lesser than 90~GeV, such as to maintain the charged scalar heavier than 100~GeV.}.
In figure~\ref{fig:1}, we show a scatter plot for the relic abundance in terms of the neutrino mass without and with coannihilation.
This figure shows that coannihilation is not determinant to have compatible models of sterile neutrinos as a viable DM candidate, but it clearly fills much more of the parameter space than the case without coannihilation. Just as an illustration we also show in figure~\ref{fig:2} the relic abundance when the Higgs mass is fixed in two cases, one for $M_H=115$~GeV and the other for $M_H=300$~GeV. In this case both graphs include coannihilation.
There is almost no quantitative difference in the relic abundance concerning the difference in Higgs mass within the range $114.4~{\mbox GeV}\leq M_H \leq 300~{\mbox GeV}$. But as we will see later, these different Higgs masses are crucial when we are concerned about direct detection experiments, mainly with respect to CDMS-II recent results~\cite{cdms}.

\begin{figure}[h]
\centering
\includegraphics[width=0.45\columnwidth]{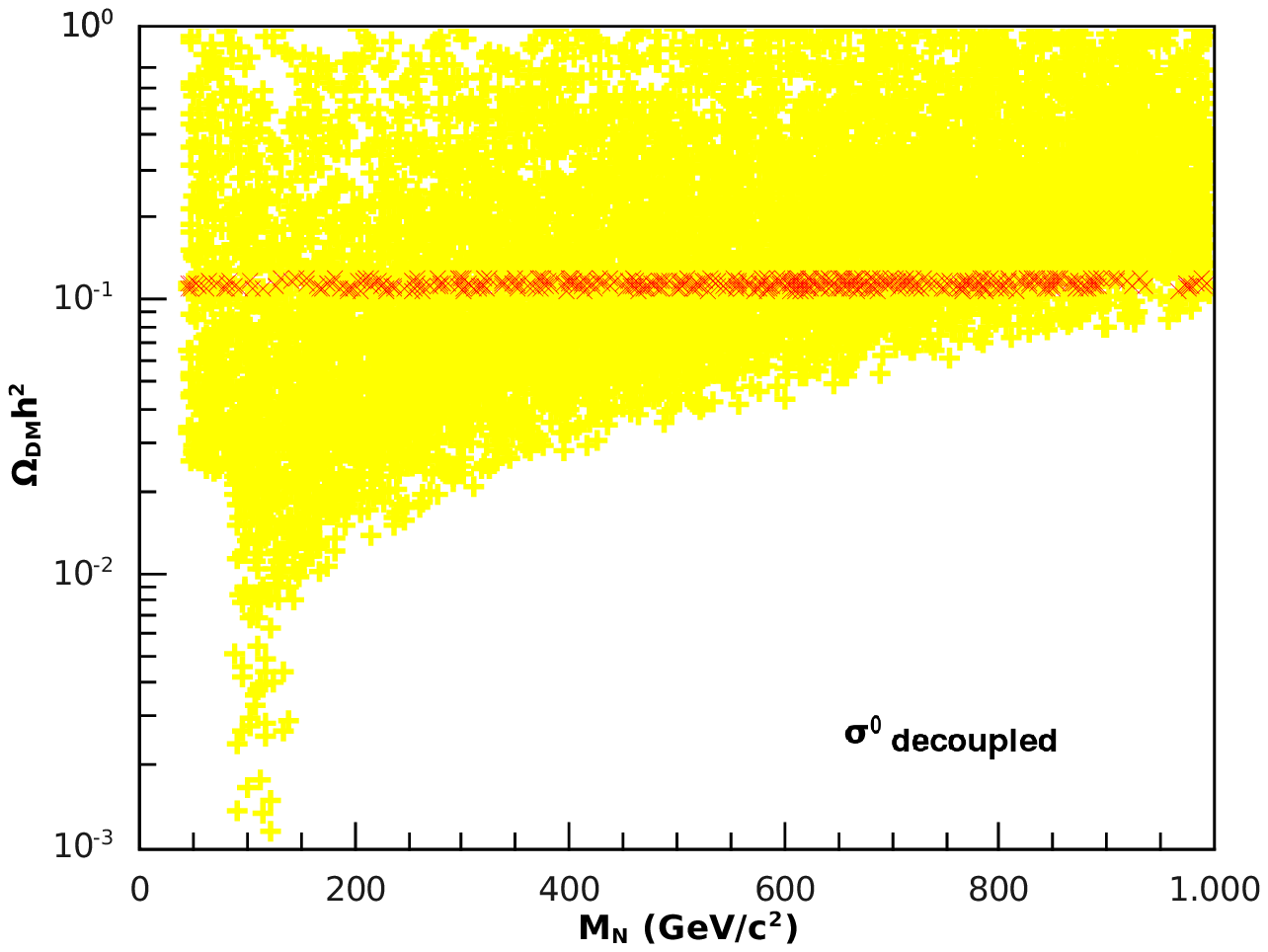}
\includegraphics[width=0.45\columnwidth]{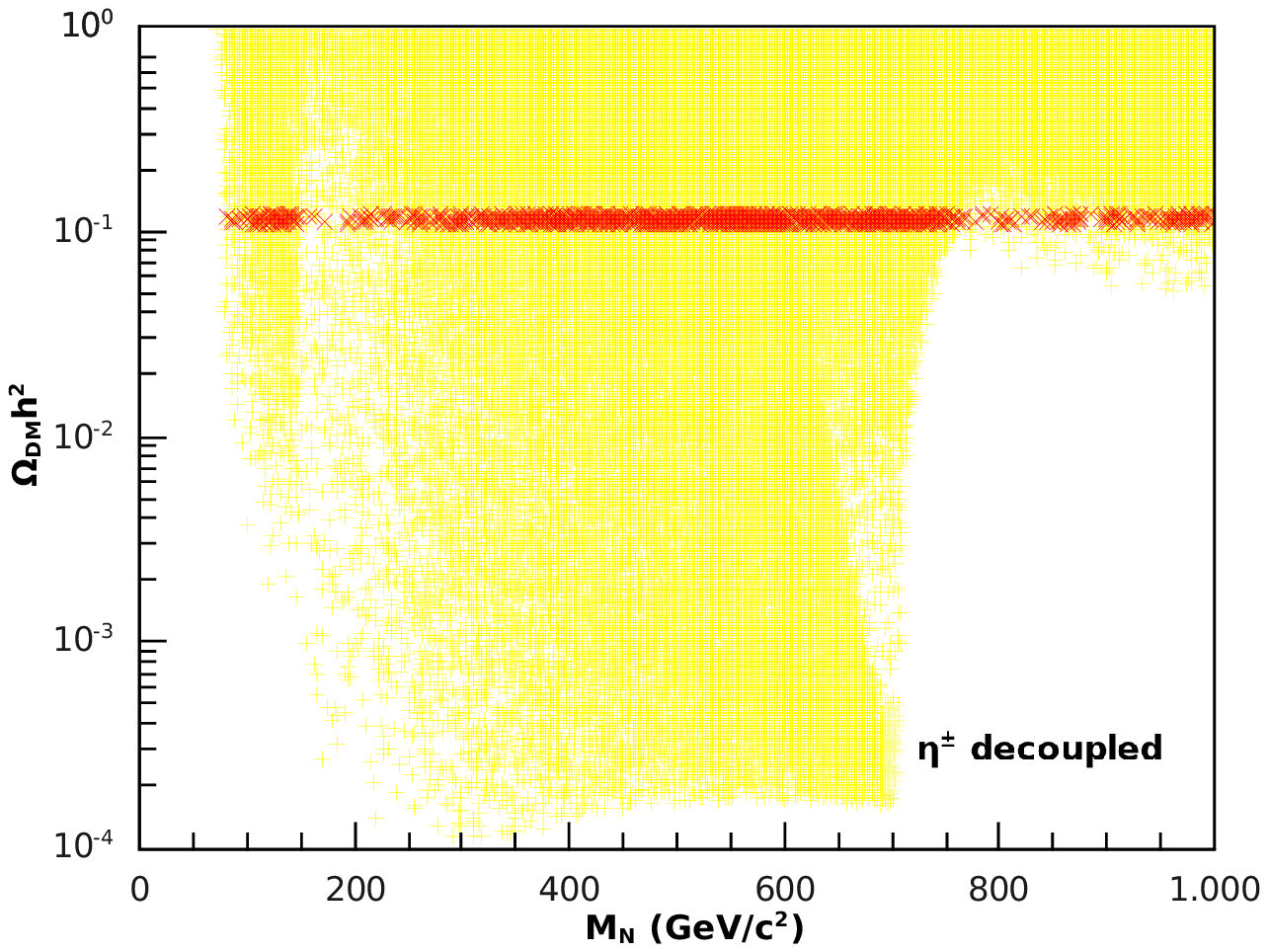}
\caption{Scatter plot for the relic abundance as a function of sterile neutrino mass. The left panel corresponds to the model without $\s^0$, while the right panel refers to the model without $\eta^\pm$. The blue cross represents points for several possible outcomes for the model considered, while the red cross represents those points in parameter space which are in agreement with WMAP 5-year run within $95\%$ CL.}
\label{fig:3}
\end{figure}

It is also interesting to consider two limit situations for this model. One in which the neutral scalar decouples from the spectrum and another one in which the charged scalar is the decoupling particle. The first situation is a peculiar one in the sense that no direct detection experiment is available to test it, since in such a case there is no mediator particle that could connect the WIMP with the nuclei. Only accelerator experiments can put a constraint on it, unless a signal is confirmed by current direct detection experiments, which would leave no room for this possibility. On the other hand, absence of any direct detection signal would favor such a possibility. We show in figure~\ref{fig:3} the results for the relic abundance in these situations for Higgs mass varying between $114.4~{\mbox GeV} \leq M_H \leq 300~{\mbox GeV}$. The first panel is for the $\s^0$-less model with coannihilation and the second panel is for the $\eta^\pm$-less model which has no coannihilation channel.
From these figures we notice that there is no restriction for the neutrino mass with the appropriate abundance, and the case with no $\eta^\pm$ is preferred only if any direct detection signal is claimed, a possibility we are going to treat in section~\ref{dd}. In any case, if for some reason, Nature chooses a scenario for the realization of this model which is not strictly the one with both, $\eta^\pm$ and $\s$, in the low energy spectrum, which means that one of them would decouple, it is interesting to check that these two situations do not describe the same physics and hence they are inequivalent and distinguishable.

We have shown that there is a lot of space to accommodate a sterile neutrino (heavy one) as a WIMP DM candidate in the model discussed here, which poses no undesirable consequences for the electroweak physics since this neutrino does not couple to the electroweak sector. It is however imperative to look for any bounds coming from direct searches of this WIMP and see the implications for the above proposal, which we do next.

\subsection{Direct detection}
\label{dd}
It is an exciting time to scrutinize the dark sector of particle physics models. Not only because we have much more information about the amount of DM in the Universe, but also because many direct detection experiments are improving the precision of cross section for DM scattering with nuclei. In this work we rely upon the most recent experimental results, XENON10~\cite{xenon} and CDMS-II~\cite{cdms}, including some prospects for future improved sensitivity of these experiments. The theoretical predictions of our model are again based on numerical computation using micrOMEGAs~\cite{micromegas}, which we present next for the cases discussed above.

\begin{figure}[h]
\centering
\includegraphics[width=0.45\columnwidth]{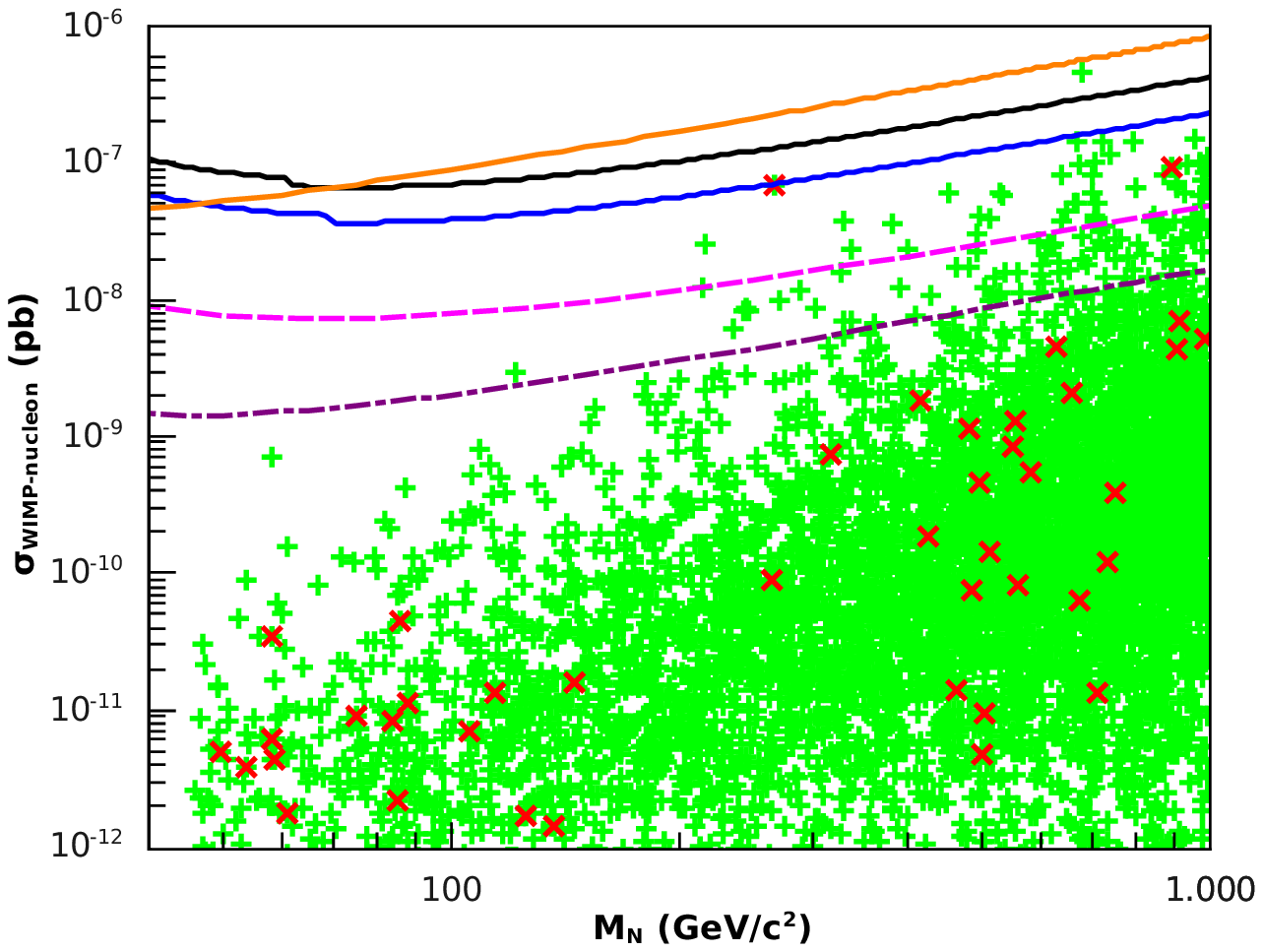}
\includegraphics[width=0.45\columnwidth]{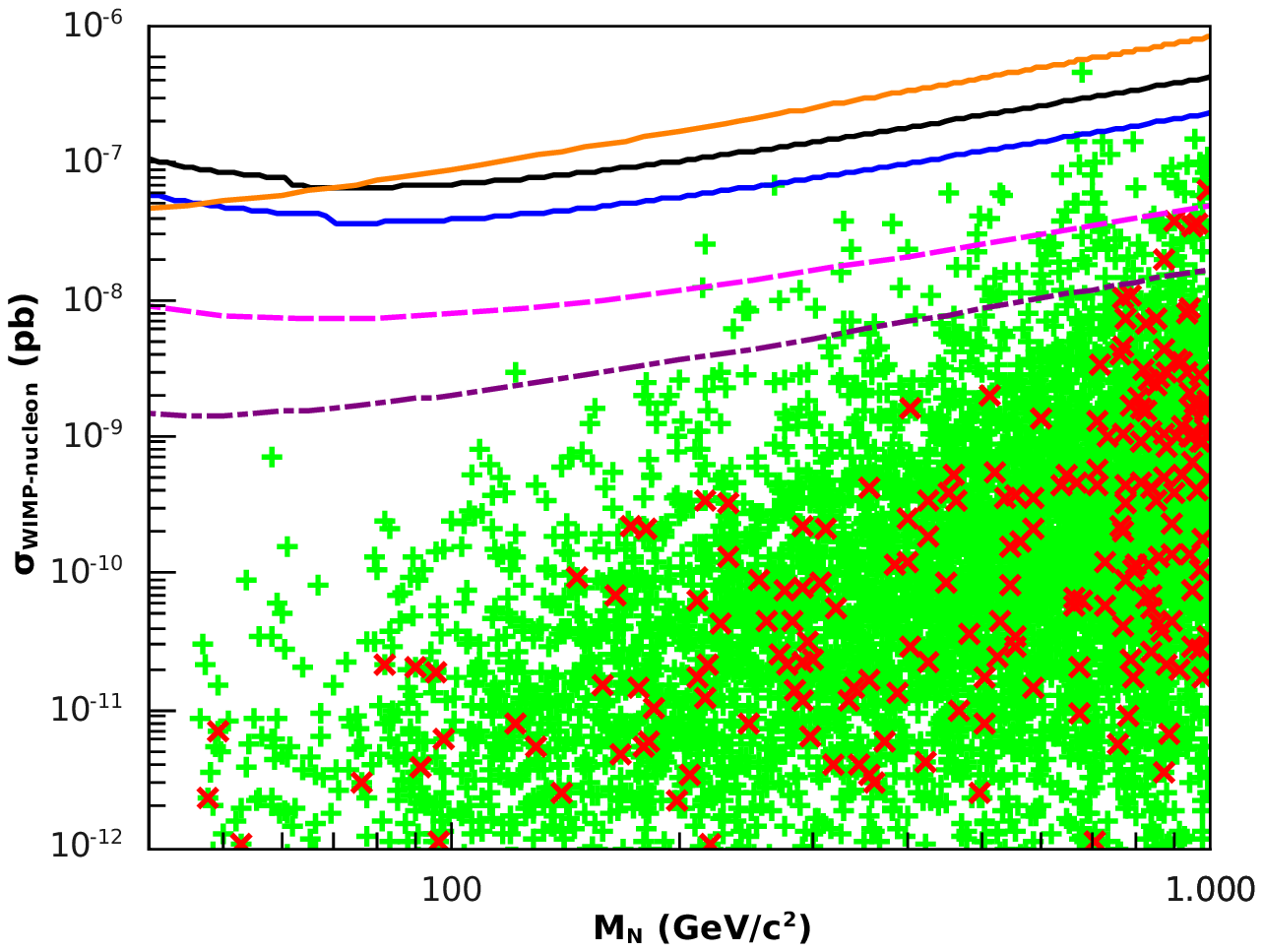}
\caption{Scatter plot for WIMP-proton cross section as a function of sterile neutrino mass. The left panel corresponds to the complete model without coannihilation, while the right panel refers to the model with coannihilation. The green cross (+) represents points for several possible outcomes for the model considered, while the red cross (x) represents those points in parameter space which are in agreement with WMAP 5-year run within $95\%$ CL. From top to bottom starting at the upper right, the full lines are experimental results, CDMS-II (black), CDMS 2004-2009 combined (blue), XENON10 (orange), the dashed line (magenta) is CDMS 2ST @ soudan prospect, the dased-dotted line (violet) is XENON100 6000kg prospect.}
\label{fig:4}
\end{figure}

\begin{figure}[h]
\centering
\includegraphics[width=0.45\columnwidth]{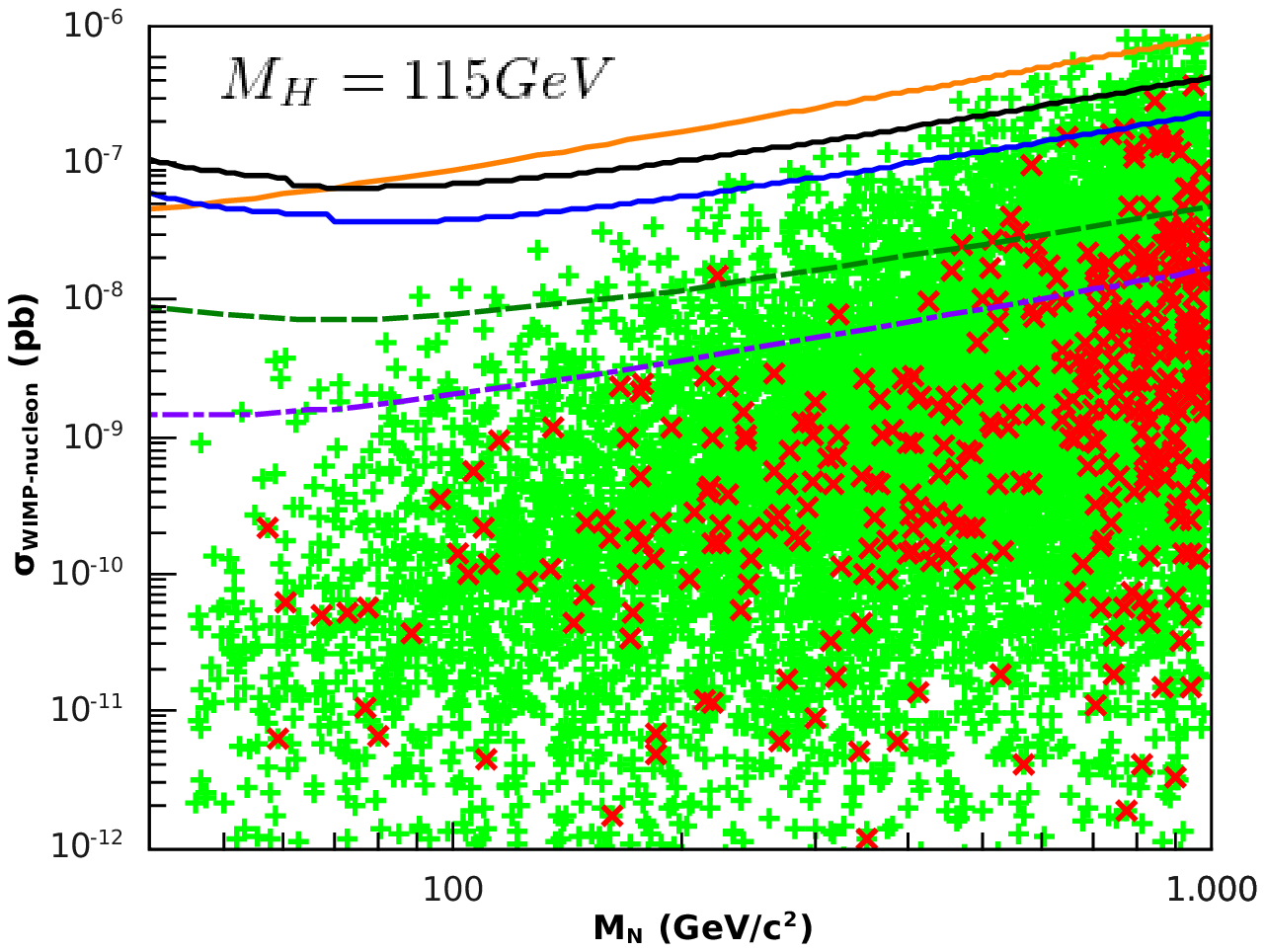}
\includegraphics[width=0.45\columnwidth]{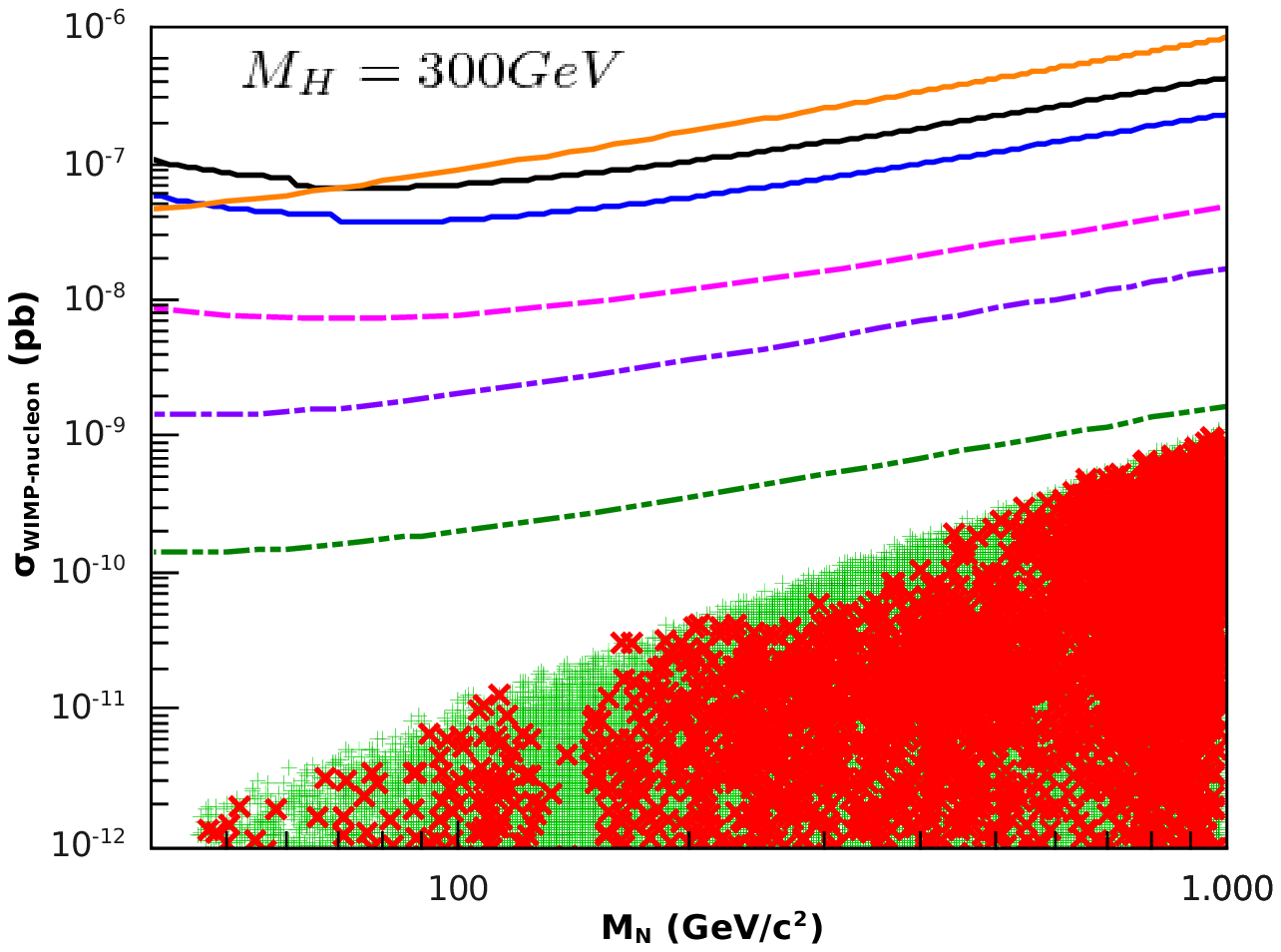}
\caption{Scatter plot for WIMP-proton cross section as a function of sterile neutrino mass (with $\eta^\pm$ and $\s$ in the spectrum). The left panel corresponds to $M_H=115$~GeV, while the right panel refers to the model with $M_H=300$~GeV. The green cross (+) represents points for several possible outcomes for the model considered, while the red cross (x) represents those points in parameter space which are in agreement with WMAP 5-year run within $95\%$ CL. From top to bottom starting at the upper right, the full lines are experimental results, CDMS-II (black), CDMS 2004-2009 combined (blue), XENON-10 (orange), the dashed line (magenta) is CDMS 2ST @ soudan prospect, the dased-dotted line (violet) is XENON100 6000kg prospect. In the right panel we included XENON-100 60000kg prospect, dashed-double dotted line (olive).}
\label{fig:5}
\end{figure}

Below, see figure~\ref{fig:4}, we show the results for the case where $\eta^\pm$ and $\sigma$ are present in the spectrum~\footnote{In all the following figures concerning DM direct detection, we used experimental data obtained from the web page in Ref.~\cite{dmtools}.}. From these plots we could say that the sterile neutrino is still to be found on direct detection experiments and there is no much qualitative difference if there is or there is not coannihilation, although coannihilation provides a much larger assessable region of the parameter space that can be tested in the near future, a feature we could infer from the results on the previous paragraph concerning relic abundance. 
We also included plots for two different Higgs mass values, $M_H=115$~GeV and $M_H=300$~GeV, see figure~\ref{fig:5}. In these plots we consider only the case where there is coannihilation since it offers a larger testable region of the parameter space.
It is noticeable that larger Higgs masses lead to less constrained parameter space, which is easy to understand since the WIMP-quark interaction in this model is due to Higgs exchange and larger Higgs masses have suppressed nuclei cross sections. We included a future prospect experiment, XENON 60000~kg in the second panel of figure~\ref{fig:5}, to stress the fact that the model with large $M_H$ is still far from being probed. When $M_H=115$~GeV, only few points above $M_N\approx 600$~GeV are excluded and most of the parameter space is in a safe region and can be assessed in the near future.

\begin{figure}[h]
\centering
\includegraphics[width=0.45\columnwidth]{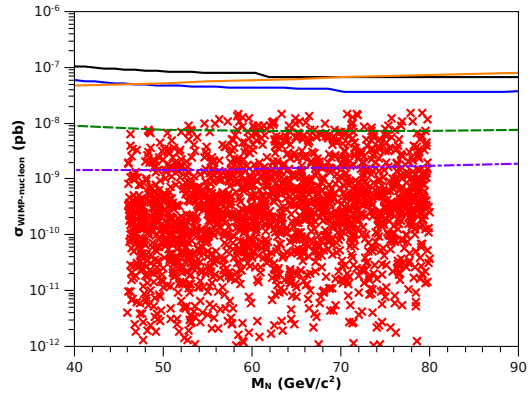}
\caption{Scatter plot for WIMP-proton cross section as a function of sterile neutrino mass. The red cross (x) represents those points in parameter space which are in agreement with WMAP 5-year run within $95\%$ CL. From top to bottom starting at the upper right, the full lines are experimental results, CDMS-II (black), CDMS 2004-2009 combined (blue), XENON10 (orange), the dashed line (olive) is CDMS 2ST @ soudan prospect, the dased-dotted line (violet) is XENON100 6000kg prospect.}
\label{fig:6}
\end{figure}

With this information at hand we make a further analysis concerning the new CDMS-II~\cite{cdms} results. It was shown in Ref.~\cite{kopp} that at 1~$\s$ there is a positive signal for detection of a dark matter recoiling from nuclei in spin independent elastic scattering. The significance is low so no detection can be claimed, however, it might be prompting for a rather light WIMP, something between $10\leq M_{DM} \leq 80$~GeV, and a cross section annihilation per nucleon ranging from about $10^{-8}~{\mbox pb}$ to $10^{-7}~{\mbox pb}$~\cite{kopp}. We have limited the range of neutrino mass within the window $46~{\mbox GeV} < M_N < 80$~GeV, and kept Higgs mass between $M_H=114.4$~GeV and $M_H=150$~GeV. Also, to get a WIMP which is in accordance with the above mentioned CDMS-II results, we lowered the value of $v_\s = 500$~GeV. Our results are shown in figure~\ref{fig:6}, where we only included those points in accordance with WMAP.
We see from this figure that there are some points in the parameter space which are marginally in agreement with a positive signal. The interesting result is that while CDMS-II seems to prefer a relatively light WIMP, our scenario where this WIMP is the sterile neutrino with couplings to singlet scalars favors a light Higgs boson and also a small scale for new Physics, which points to an exciting possibility for the forthcoming tests at LHC.

\begin{figure}[h]
\centering
\includegraphics[width=0.45\columnwidth]{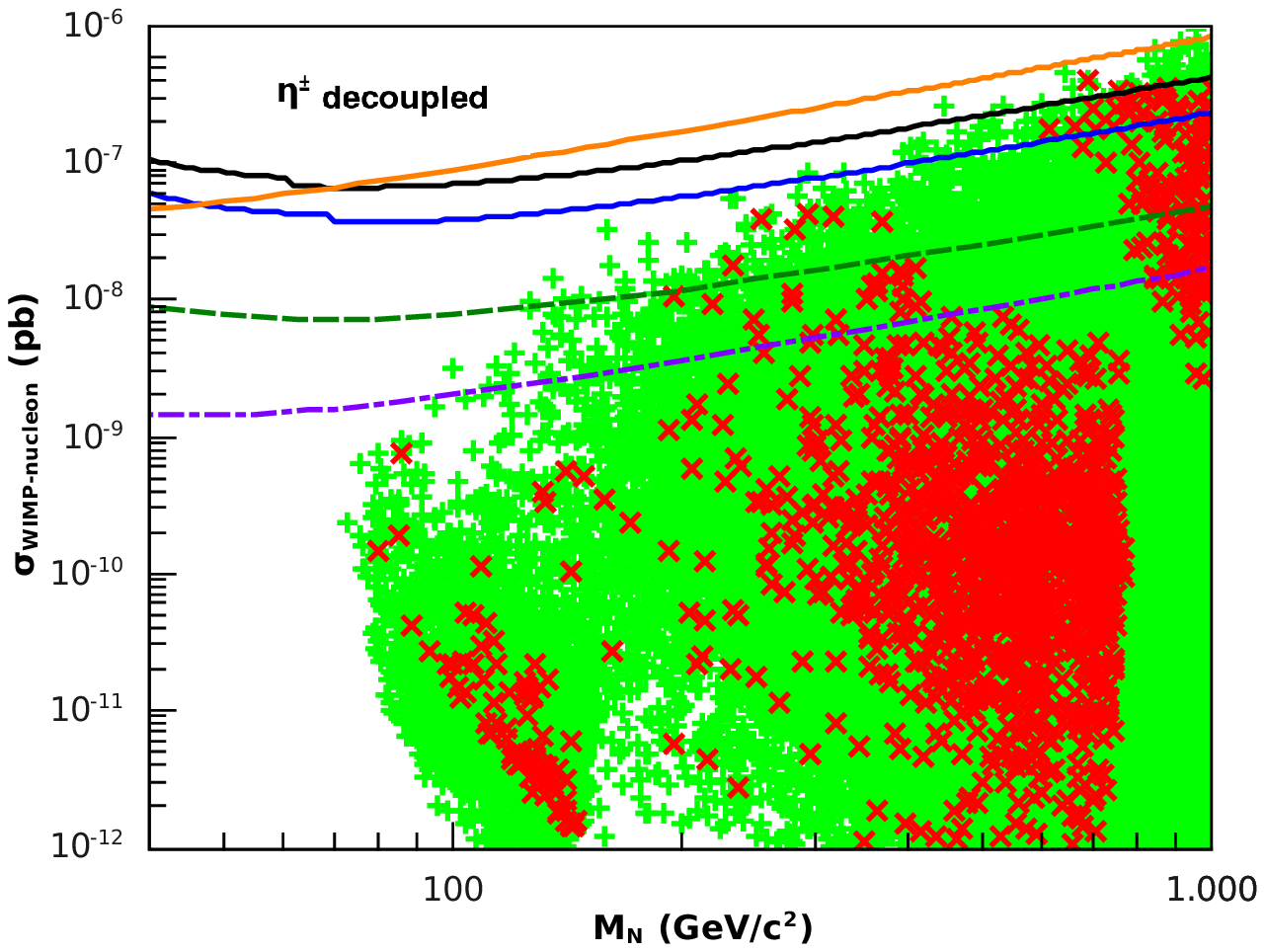}
\includegraphics[width=0.45\columnwidth]{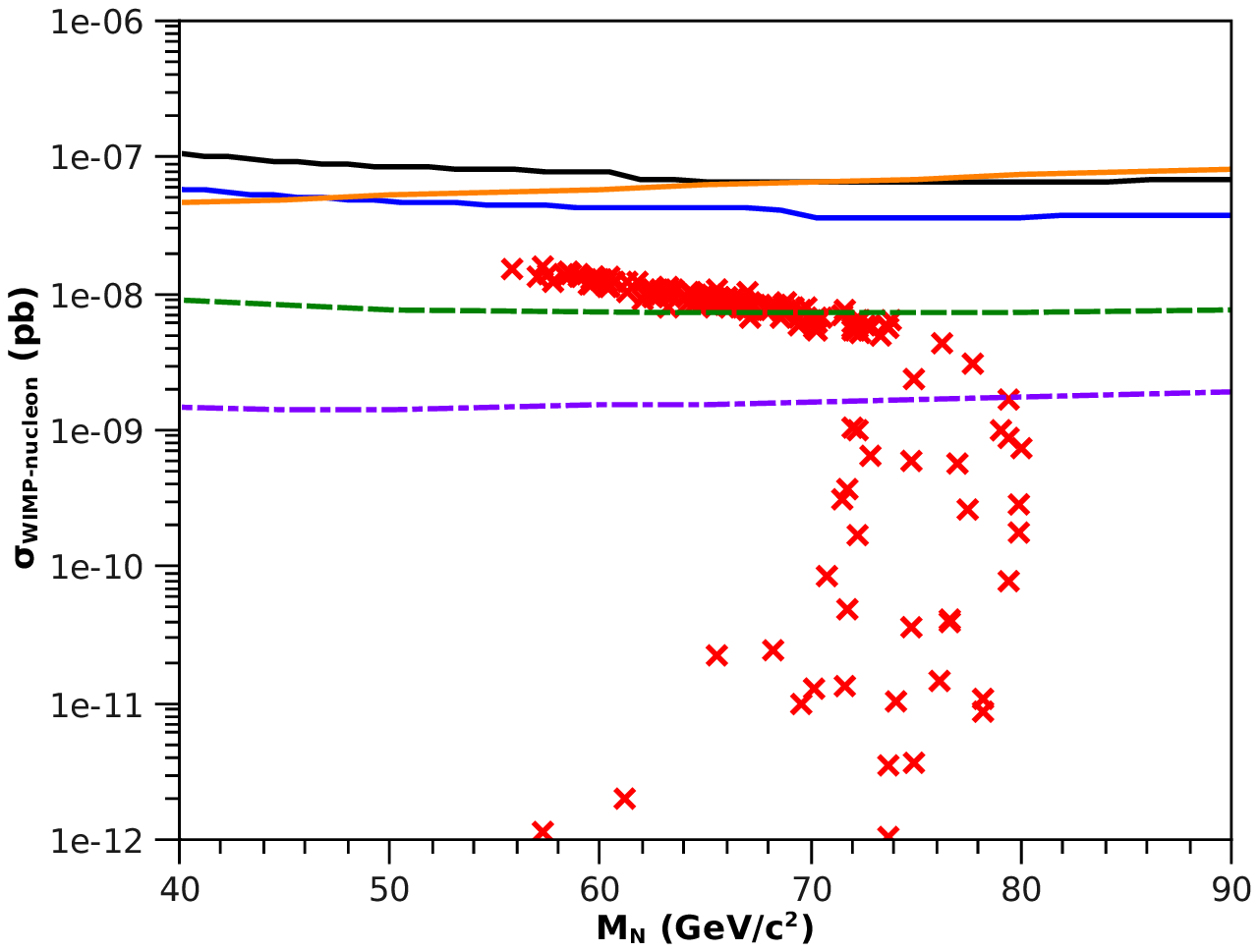}
\caption{Scatter plot for WIMP-proton cross section as a function of sterile neutrino mass (with only  $\s$ in the spectrum). The green cross (+) represents points for several possible outcomes for the model considered, while the red cross (x) represents those points in parameter space which are in agreement with WMAP 5-year run within $95\%$ CL. From top to bottom starting at the upper right, the full lines are experimental results, CDMS-II (black), CDMS 2004-2009 combined (blue), XENON10 (orange), the dashed line (olive) is CDMS 2ST @ soudan prospect, the dased-dotted line (violet) is XENON100 6000kg prospect. In the right panel we included XENON100 60000kg prospect, dashed-double dotted line.}
\label{fig:7}
\end{figure}

Finally, the case of $\s$ decoupling, as we discussed before, provides no direct detection signal since it involves only the exchange of a new charged scalar bilepton, with no coupling to quarks. This situation recovers that of Ref.~\cite{ma}. By the other side, if the charged scalar bilepton has a large mass and decouples, we can check the viability of the model in the light of direct detection experiments. This situation recovers the sterile neutrino WIMP of Ref.~\cite{GHSV} and the results are shown in the first panel of figure~\ref{fig:7} where we can see that there is good agreement with experiments up to a mass of 1~TeV for some region of the parameter space.

We also included the direct detection results for this scenario, with $\eta^\pm$ decoupled, for small neutrino masses aiming to see the constraints from CDMS-II experiment. In the second panel of figure~\ref{fig:7} we can see that there are points in the parameter space in compliance with a positive CDMS-II signal for small neutrino mass, again implying a light Higgs boson, $115~{\mbox GeV}< M_H < 150~{\mbox GeV}$, and low energy new Physics, $v_\s \approx 500$~GeV. As was already pointed in Ref.~\cite{GHSV}, a collider search for this neutrino through Higgs boson decay, $H\to N+N$, can be enhanced in comparison to standard Yukawa interactions for light fermions since this coupling does not need to be small. In our case it is given by Eq.~(\ref{coupl}), and in view of the values we adopted for the parameters in the Eqs.~(\ref{lagr}) and (\ref{pot}), $\l_1=\l_5=\l_\s\approx 0.1$, $v = 246$~GeV, and $v_\s = 500$~GeV, this coupling is about $\l_{N-N-H}\approx 0.02$, which compares with the $SU_L(2)$ gauge coupling. 

\section{Conclusions}
\label{sec3}

We have defined a scenario for Physics beyond SM where a singlet right-handed neutrino is added together with a charged and a neutral singlet scalars. This scenario may represent some low energy regime of gauge or Higgs extensions of SM~\cite{GHSV,ma,331rh}. The right-handed neutrino proves to be a stable sterile neutrino if a discrete symmetry is assumed, and thus represent a good candidate for the DM in the Universe. We have probed the parameter space of this model and found that there are regions where the model is compatible with recent results of WMAP-5, allowing this sterile neutrino to represent the main component of DM. Also, we have tested this model under direct DM detection experimental bounds and checked that it is mostly safe for a range of neutrino masses between $M_N=46$~GeV to $M_N=1$~TeV when Higss mass is $114.4~{\mbox GeV} < M_H < 300$~GeV. However, we noticed that lowest Higgs masses provides more interesting results in the sense that they can be probed in such DM detection experiments sooner than higher ones. Finally, when restricting to low neutrino masses in order to be in the mass region favored by CDMS-II, we observed that only the lowest Higgs mass together with a small scale for new Physics, in our case this is established by the neutral singlet scalar VEV, $v_\s\approx 500$~GeV, are marginally in agreement with a $1\s$ positive signal from CDMS-II~\cite{cdms,kopp}. Although it is too soon to claim detection with so low statistics, it maybe that CDMS-II is pointing to an intriguing interplay among neutrino physics, DM problem and Higgs search.

\section*{Acknowledgments}

We thank J. K. Mizukoshi for useful discussions. We also would like to thank A. Semenov for valuable information about LanHEP and CalcHEP and A. Pukhov for his promptly help concerning micrOMEGAs package. This research was supported by the Conselho Nacional de Desenvolvimento Cient\'{\i}fico e Tecnol\'ogico (CNPq) (FSQ, CASP and PSRS) and Coordena\c{c}\~ao de aperfei\c{c}oamento de Pessoal de N\'{\i}vel Superior (CAPES) (FSQ). 

{\section*{References}}
\bibliographystyle{JHEP}

\end{document}